\documentstyle[prd,twocolumn,aps]{revtex}

\newcommand{\be}{\begin{equation}}
\newcommand{\ee}{\end{equation}}

\begin{document}
\twocolumn[\hsize\textwidth\columnwidth\hsize\csname @twocolumnfalse\endcsname
\draft
\title{Teleparallel Spin Connection}

\author{V. C. de Andrade}
\address{ D\'epartement d'Astrophysique Relativiste et de Cosmologie \\
Centre National de la Recherche Scientific (UMR 8629) \\
Observatoire de Paris, 92195 Meudon Cedex, France}

\author{L. C. T. Guillen and J. G. Pereira}
\address{Instituto de F\'{\i}sica Te\'orica,
Universidade Estadual Paulista \\
Rua Pamplona 145, 01405-900 \, S\~ao Paulo SP, Brazil}
\date{\today}
\maketitle

\begin{abstract}

A new expression for the spin connection of teleparallel gravity is proposed, given by
{\it minus} the contorsion tensor plus a {\em zero} connection. The corresponding minimal
coupling is covariant under local Lorentz transformation, and equivalent to the minimal coupling
prescription of general relativity. With this coupling prescription, therefore, teleparallel
gravity turns out to be fully equivalent to general relativity, even in the presence of spinor
fields.

\end{abstract}

\vskip1pc]

In the absence of spinor fields, the equivalence of teleparallel gravity~\cite{hehl} with general
relativity is a well established subject~\cite{paper1}. In the presence of spinor fields,
however, it is widely believed that this equivalence is spoiled~\cite{trans}. The reason for
this is that, in teleparallel gravity, the spin connection is believed to vanish. The  basic
purpose of this report is to show that this conclusion is not appropriate, and that it comes
from a particular choice of the teleparallel spin connection. A new expression for the spin
connection is then proposed, according to which teleparallel gravity becomes much more
consistent and fully equivalent to general relativity, even in the presence of spinor fields.

We use the Greek alphabet $(\mu, \nu, \rho, \dots = 0,1,2,3)$ to denote indices related to
spacetime (base space), and the Latin alphabet $(a,b,c, \dots = 0,1,2,3)$ to denote indices
related to the tangent space (fiber), assumed to be a Minkowski spacetime. The spin connection
$\stackrel{\circ}{A}_{\mu}$ can thus be written as
\be
\stackrel{\circ}{A}_{\mu} = \frac{1}{2} \,
{\stackrel{\circ}{A}}{}^{a b}{}_{\mu} \, S_{a b} \; ,
\label{spinco}
\ee
where $S_{a b}$ is an element of the Lorentz Lie algebra written in an appropriate
representation.

In general relativity, the spin connection is~\cite{dirac}
\be
{\stackrel{\circ}{A}}{}^{a}{}_{b \nu} = h^{a}{}_{\rho}
{\stackrel{\circ}{\Gamma}}{}^{\rho}{}_{\mu \nu} \, h_{b}{}^{\mu} +
h^{a}{}_{\rho} \partial_\nu h_{b}{}^{\rho} \; ,
\label{gra}
\ee
where $h^a{}_{\mu}$ is a tetrad field, and ${\stackrel{\circ}{\Gamma}}{}^{\rho}{}_{\mu \nu}$ is
the Levi--Civita connection. In order to obtain the teleparallel spin connection, inspired by
the definition (\ref{gra}), it is usual to start by making the following attempt,
\be
A^{a}{}_{b \nu} = h^{a}{}_{\rho} \, \Gamma^{\rho}{}_{\mu \nu} \, 
h_{b}{}^{\mu} +
h^{a}{}_{\rho} \partial_\nu h_{b}{}^{\rho} \; ,
\label{telecon}  
\ee
with $\Gamma^{\rho}{}_{\mu \nu}$ the Weitzenb\"ock connection. However, as a consequence of the
absolute parallelism condition~\cite{paper1}, we have $A^{a}{}_{b \nu}=0$. This does not mean
that in teleparallel gravity the spin connection vanishes. It only means that we have not made
an appropriate guess.

Let us then adopt a different procedure to look for the teleparallel spin connection. Our basic
guideline will be to find a coupling prescription that results equivalent to the coupling
prescription of general relativity. This can be achieved by taking the relation between the
Levi--Civita and the Weitzenb\"ock connections~\cite{prl},
\be
{\stackrel{\circ}{\Gamma}}{}^{\rho}{}_{\mu \nu} = - K^{\rho}{}_{\mu \nu} +
\Gamma^{\rho}{}_{\mu \nu} \; ,
\ee
with $K^{\rho}{}_{\mu \nu}$ the contorsion tensor, and rewriting it in the tetrad basis. The
result is
\be
{\stackrel{\circ}{A}}{}^{a}{}_{b \mu} = - K^{a}{}_{b \mu} + 0 \; ,
\label{tsc}
\ee
where we have already used the fact that $A^{a}{}_{b \mu}=0$. Notice that the {\em zero
connection} appearing in Eq.(\ref{tsc}) is crucial in the sense that it is the responsible for
making the right hand-side a true connection. Therefore, based on these considerations, we can
say that the teleparallel spin connection $\omega^{a}{}_{b \mu}$ is given by {\em minus} the
contorsion tensor plus a zero--connection:
\be
\omega^{a}{}_{b \mu} = - K^{a}{}_{b \mu} + 0 \; .
\ee
Accordingly, the teleparallel Fock-Ivanenko derivative operator is to be written in the form
\be
{\mathcal D}_\mu = \partial_\mu - \frac{i}{2} \,  
\omega^{a}{}_{b \mu} \, S_a{}^b \; .
\label{tfi}
\ee
It is important to remark that, differently from ${\stackrel{\circ}{A}}{}^{a}{}_{b \mu}$ which
depends on the Levi-Civita connection only, $\omega^{a}{}_{b \mu}$ depends on the Weitzenb\"ock
connection only.

The covariant derivative (\ref{tfi}) presents all necessary properties to be considered as
yielding the fundamental coupling prescription in teleparallel gravity. It transforms
covariantly under local Lorentz transformations, it reduces to the teleparallel version of the
Levi--Civita covariant derivative when applied to tensor fields~\cite{vector}, and it turns out
to be completely equivalent to the usual minimal coupling prescription of general relativity,
even in the presence of spinor fields. Therefore, we can conclude that $\omega^{a}{}_{b \mu}$
presents all necessary properties to be considered as the spin connection of teleparallel
gravity. The consistency of this result can also be checked by noticing that, under a local
Lorentz transformation with parameters
$\epsilon^a{}_b$, the spin connection $\omega^{a}{}_{b \mu}$ changes according to
\be
\delta \omega^{a}{}_{b \mu} = {\mathcal D}_\mu \epsilon^a{}_b \; ,
\label{ktrans1}
\ee
which is the standard connection transformation of gauge theories~\cite{ramond}.

\section*{Acknowledgments}

The authors are indebted to R. Aldrovandi and C. M. Zhang for valuable comments. They would
like also to thank FAPESP-Brazil, CAPES-Brazil and CNPq-Brazil for financial support.

\end{document}